\documentclass[12pt]{article}

\usepackage{psfig}

\usepackage[dvips]{color}

\newcommand{\Black}{\color [rgb]{0,0,0}}

\newcommand{\Brown}{\color [rgb]{0.4,0.1,0.1}}

\def\TL{\hfil$\displaystyle{##}$}
\def\TR{$\displaystyle{{}##}$\hfil}

\def\TT{\hbox{##}}
\def\seqalign#1#2{\vcenter{\openup1\jot
  \halign{\strut #1\cr #2 \cr}}}

\def\fixit#1{}

\def\mop#1{\mathop{\rm #1}\nolimits}

\def\Re{\mop{Re}}

\def\Vol{\mop{Vol}}

\def\overleftrightarrow#1{\vbox{\ialign{##\crcr
     $\leftrightarrow$\crcr\noalign{\kern-0pt\nointerlineskip}
     $\hfil\displaystyle{#1}\hfil$\crcr}}}

\def\lsim{\mathrel{\mathstrut\smash{\ooalign{\raise2.5pt\hbox{$<$}\cr\lower2.5pt\hbox{$\sim$}}}}}
\def\gsim{\mathrel{\mathstrut\smash{\ooalign{\raise2.5pt\hbox{$>$}\cr\lower2.5pt\hbox{$\sim$}}}}}

\def\sqr#1#2{{\vcenter{\vbox{\hrule height.#2pt
         \hbox{\vrule width.#2pt height#1pt \kern#1pt
            \vrule width.#2pt}
         \hrule height.#2pt}}}}

\def\href#1#2{#2}

\def\lbldef#1#2{\expandafter\gdef\csname #1\endcsname {#2}}
\def\eqn#1#2{\lbldef{#1}{(\ref{#1})}%
\begin{equation} \eqalign{#2} \label{#1} \end{equation}}
\def\eqalign#1{\vcenter{\openup1\jot
    \halign{\strut\span\TL & \span\TR\cr #1 \cr
   }}}

\begin{document}
\pagestyle{plain}
\setcounter{page}{1}
\begin{titlepage}

\begin{flushright}
PUPT-2084 \\
hep-th/0305099
\end{flushright}
\vfil

\begin{center}
{\huge String production \\[10pt]
at the level of effective field theory}
\end{center}

\vfil
\begin{center}
{\large Steven S. Gubser}
\end{center}

$$\seqalign{\span\TL & \span\TT}{
& Joseph Henry Laboratories, Princeton University, Princeton, NJ 08544
}$$
\vfil

\begin{center}
{\large Abstract}
\end{center}

\noindent
 Pair creation of strings in time-dependent backgrounds is studied from
an effective field theory viewpoint, and some possible cosmological
applications are discussed.  Simple estimates suggest that excited
strings may have played a significant role in preheating, if the
string tension as measured in four-dimensional Einstein frame falls a
couple of orders of magnitude below the four-dimensional Planck
scale.

\vfil
\begin{flushleft}
May 2003
\end{flushleft}
\end{titlepage}
\newpage

\section{Introduction}

In curing the non-renormalizability of gravity, string theory
introduces a large number of heavy states: the excited states of a
string.  The number of these states increases roughly as an
exponential of their energy, in contrast to field theories, where the
number of states rises as a power of the energy.  This exponential
increase is often described as a Hagedorn density of states.  The
extra states in string theory are conventionally assumed to lie near
the Planck scale.  Our grasp of the non-perturbative dynamics of
string theory is far from complete, but one well-motivated conjecture
is that interactions vastly reduce the degrees of freedom of the
theory, so that the number of available states scales
``holographically'' as the area of a system rather than its volume.
Disentangling this ``more is less'' paradox, elucidating the true
degrees of freedom of string theory in a non-perturbative regime, and
reconciling the renormalization group with holography may all be
necessary steps before we can give a fully satisfactory string
theoretic account of the very early universe.  We are a long way from
achieving this, but certainly there is ample reason to believe that
theories with extended objects are needed to properly formulate
quantum gravity, and that such theories often have a Hagedorn density
of states over an energy range encompassing a great many states.
Armed with no more than this, we would like to inquire what the
possible consequences are for cosmology---in particular, for
situations where the massive states may be produced through the usual
ambiguity of the vacuum state in backgrounds with time evolution.  We
will start with an overview of pair production and the steepest
descent method for estimating occupation numbers.  Then we will move
on to a possible application to the theory of preheating.

In regimes of parameter space where a spacetime description gives a
good approximation of string dynamics, the on-shell constraint for a
given string state boils down to a second order ordinary differential
equation in time:
 \eqn{ChiEOM}{
  \ddot{\chi} + \omega(t)^2 \chi = 0 \,,
 }
where $\chi(t)$ is the wave function for the state in question.  This
is also the equation for a mode of a scalar field, possibly rescaled
by some time-dependent factor to eliminate a $\dot\chi$ term.  The
function $\omega(t)$ must be determined for any given background and
string state.  Quantizing strings in a general spacetime background is
fraught with difficulties because the spacetime equations of motion
are already encoded in the conditions for conformal invariance on the
worldsheet, and we seem to be missing some aspects of string dynamics
which give rise to the weakly curved spacetime we observe.  If we can
quantize strings in a particular background, then $\omega(t)$ will be
determined, and it will include contributions from the momentum of the
string and its excitation state.  In the following treatment, we will
simply assume that the spectrum is known, and that it includes the
familiar Hagedorn density of states.  We are interested in
establishing conditions under which the total string production is
finite (without invoking a cutoff on the Hagedorn spectrum), and in
estimating the total rate of string production when it is dominated by
highly excited strings.  For this reason, we take $\omega(t)^2$ to be
large.

To speak meaningfully about string pair production, it's necessary at
least to have an asymptotic ``out'' region where $\omega$ is slowly
varying, in the sense that time derivatives of $\omega$ are much
smaller than the power of $\omega$ with the same dimension.  When
there is such a region, one can compare the infinite order out
adiabatic vacuum to the actual quantum state to determine occupation
numbers for a given string state.  What the actual quantum state is
can be subtle, but if there is an asymptotic ``in'' region where again
$\omega$ is slowly varying, one can follow the standard approach of
letting the actual quantum state correspond to the in vacuum.  Let us
consider this optimal situation first and further assume that $t$ runs
from $-\infty$ to $+\infty$.

Two widespread analytical techniques for extracting approximate pair
production rates from \ChiEOM\ are steepest descent contours,
applicable when particle occupation numbers are small, and parametric
resonance, applicable when $\omega(t)^2$ is an oscillatory function of
time.  We will be mainly interested in the former and will develop it
in section~\ref{Steepest}.  We will then give an application to the
theory of preheating in section~\ref{Preheating}, where conventionally
one needs parametric resonance to understand the physics---but as we
shall see, the steepest descent method is still suitable for
discussing production of excited string states.  The conclusion will
be that there is a plausible regime of parameters where strings of
some type played a significant roll in preheating, but that
over-production of super-heavy dark matter is a potential problem.

After this work was complete, I learned of \cite{lmCreate},\footnote{I
thank E.~Martinec for bringing this paper to my attention.} which
overlaps significantly with the methodology developed in
section~\ref{Steepest}.

\section{The steepest descent method}
\label{Steepest}

The steepest descent method is based on early work by
\cite{BrezinItzykson} and has been developed by various authors.  Our
treatment will to an extent parallel that of \cite{ChungHeavy}.  The
key assumption is that the occupation number $|\beta|^2$ for a given
mode is always much less than $1$.  Here, $\beta$ is a Bogliubov
coefficient for comparing the in and out vacua.\footnote{In the
standard analogy to one-dimensional scattering, where time is mapped
to position and \ChiEOM\ is regarded as a time-independent Schrodinger
equation with potential $-\omega^2(t)$, $\beta$ is roughly the
reflection amplitude.}  Setting
 \eqn{IntroduceAlphaBeta}{
  \chi(t) = {\alpha(t) \over \sqrt{2\omega(t)}}
    e^{-i \int^t du \, \omega(u)} + 
   {\beta(t) \over \sqrt{2\omega(t)}}
    e^{i \int^t du \, \omega(u)} \,,
 }
with the requirement $|\alpha(t)|^2-|\beta(t)|^2=1$, one may recast the
equation \ChiEOM\ as
 \eqn{ABEOM}{
  \dot\alpha(t) = {\dot\omega \over 2\omega} 
   e^{2i\int^t du \, \omega(\tilde(t))} \beta(t) \qquad
  \dot\beta(t) = {\dot\omega \over 2\omega} 
   e^{-2i\int^t du \, \omega(\tilde(t))} \alpha(t) \,.
 }
Using $\beta(t) \ll 1$ and $\alpha(t) \approx 1$, one quickly arrives
at the general formula for $\beta = \beta(\infty)$:
 \eqn{MasterBeta}{
  \beta \approx \int_{-\infty}^\infty dt \, {\dot\omega \over 2\omega}
   \exp\left( -2i \int^t du \, \omega(u) \right) \,.
 }
Assuming spacetime is weakly curved in the out region, that $|g_{tt}|
\to 1$ there, and that $\omega$ approaches some constant
$\omega_\infty$ for any given string state, the density of strings
states rises roughly as $e^{\omega_\infty/T_H}$, where $T_H$ is the
Hagedorn temperature,\footnote{Unusual circumstances might invalidate
this description of the Hagedorn density, for instance the boundary
operator in \cite{senRoll,stromingerEtAl,maldacenaEtAl} that gives a
given open string state a mass that grows exponentially with time.  In
this particular circumstance, $\omega_\infty$ in the discussion above
could be replaced by $\omega = \omega_0$ evaluated at the
time-symmetric point of the full s-brane solution.}  
 \eqn{THdef}{
  T_H = {1 \over 2\pi \sqrt{\alpha' c_\perp/6}}
 }
for closed strings, where the string tension is $\tau =
1/(2\pi\alpha')$ and $c_\perp$ is the central charge of the transverse
degrees of freedom ($c_\perp=12$ for the type~II superstring), assumed
here to be the same in holomorphic and anti-holomorphic sectors.  This
exponential behavior is modified by a power law that depends on the
string theory in question as well as the dimensionality of non-compact
spacetime.  Assuming $|\beta|^2$ can be approximated by some function
of $\omega_\infty$, the total number of strings produced may be very
roughly estimated to be
 \eqn{TotNum}{
  N_{\rm tot} \sim \int^\infty d\omega_\infty \, 
   |\beta|^2 e^{\omega_\infty/T_H} \,.
 }
In s-brane decay (with $\omega_\infty$ replaced by $\omega_0$, as
explained in the footnote), the exponential growth of states is
exactly canceled by exponential suppression of $\beta$ at large
$\omega$, leaving a power law behavior that may or may not converge,
depending on the dimension
\cite{AndyCreate,stromingerEtAl,maldacenaEtAl}.  Our estimates will
focus on the exponential behavior in more generic circumstances.

A well-known formal tool for exploring the high-energy properties of
\ChiEOM\ is to rescale $t \to t/\sigma$, where $\sigma$ is a
dimensionless constant.  A naive expectation, which will turn out to be
close enough to the truth in some interesting cases, is that the
on-shell condition for highly excited string states is given by the
rescaled equation,
 \eqn{RescaledEOM}{
  \ddot\chi + \sigma^2 \omega(t)^2 \chi = 0 \,,
 }
with $\sigma^2$ set equal to the excitation level $N$, and with
$\omega$ remaining nearly the same for different string states.  More
precisely, the assumption is that {\it without} any rescaling of time,
highly excited string states have $\omega \approx \sigma \bar\omega$
with $\sigma = \sqrt{N}$ and $\bar\omega$ nearly independent of the
string state.  Then \RescaledEOM\ is correct, with $\omega \to
\bar\omega$, though for a reason orthogonal to time rescaling.
Granting such a setup, the total number of strings produced is
 \eqn{TotNumRescaled}{
  N_{\rm tot} &\sim \int^\infty d\sigma \, |\beta(\sigma)|^2 
   e^{\sigma \bar{\omega}_\infty / T_H}  \cr
  \beta(\sigma) &\approx \int_{-\infty}^\infty dt \, 
   {\dot{\bar\omega} \over 2\bar\omega} 
   \exp\left( -2i \sigma \int^t du \, \bar\omega(u) \right) \,.
 }
In \TotNumRescaled, $\bar\omega_\infty/T_H$ is a fixed number,
independent of the string state.  Certainly \TotNumRescaled\ has been
arrived at through a series of assumptions that are far from
self-evident.  However it has some nice consequences that we believe
are more general.  First of all, $\bar\omega(t)^2$ must infinitely
differentiable on the real axis in order to avoid producing an
infinite number of strings.  The same arguments used in \cite{bd} to
show that occupation numbers in the adiabatic vacuum of order $A$
scale as $\omega^{-(A+1)}$ can be adapted to show that $\beta$ scales
as $\sigma^{-n}$ for large $\sigma$ when there is a discontinuity in
the $n$-th derivative of $\bar\omega(t)^2$.  Probably it is also
necessary for $\bar\omega(t)^2$ to be analytic for real $t$---we will
have more to say about this point later.  It is intriguing that
analyticity is also characteristic of the statistical mechanics of
systems with a finite number of degrees of freedom, hinting once
again that the number of degrees of freedom for gravitationally
coupled strings is finite.

To evaluate the expression for $\beta$ in \MasterBeta, a technique
based on contour integration and steepest descent was developed in
\cite{ChungHeavy}.  At least in simple circumstances, the
singularities of the outer integrand are poles and branch points,
occurring in the complex plane where $\omega(t)^2 = 0$ or $\infty$.
These singularities are distributed symmetrically on either side of
the real axis because $\omega(t)^2$ is real-valued for real arguments.
Singularities arising from simple zeroes of $\omega(t)^2$ were treated
in detail in \cite{ChungHeavy}, and through a steepest method the
following estimate was obtained:
 \eqn{SteepestApprox}{
  \beta \approx {i\pi \over 3} 
   \exp\left( -2i \int_{-\infty}^r dt \, \omega(t) \right)
   \exp\left( -2i \int_r^{t_*} dt \, \omega(t) \right) \,,
 }
where $t_* = r - i\mu$ is the location of a zero of $\omega(t)^2$ in
the lower half plane, and $r$ and $\mu$ are real.  If there are
several zeroes, one gets a sum of terms of the type appearing on the
right hand side of \SteepestApprox.  Assuming that one zero dominates,
we may roughly estimate
 \eqn{MuApprox}{
  |\beta|^2 \approx \left( {\pi \over 3} \right)^2 
   e^{-\pi \mu \omega(r)} \,.
 }
The factor of $\pi$ in the exponent arises from approximating $\Re
\omega(t)$ along the line between $r$ and $t_*$ by an elliptical arc
which passes through zero at $t_*$ and has its apex at $r$.  A better
estimate may be obtained if more information is available for the
function $\omega(t)^2$.  With the estimate \MuApprox\ in hand, we can
return to \TotNumRescaled\ and obtain
 \eqn{TooMany}{
  N_{\rm tot} \sim \int^\infty d\sigma \, 
   e^{\sigma(-\pi\mu\bar\omega(r)+\bar\omega_\infty/T_H)} \,.
 }
Evidently, this converges provided the exponent is negative.  If we
wished to compute the total energy of the strings created, it would
only alter the power law prefactor in the integrand of \TooMany, and
the criterion for convergence would still be that the exponent is
negative.

We derived \TooMany\ on the understanding that $|g_{tt}| \to 1$ as $t
\to \infty$.  If this is not so, but there is still an appropriate
asymptotic out region, then we need only replace $\bar\omega_\infty$
by $\sqrt{g^{tt}_\infty} \bar\omega_\infty$ (defined as a limit).
Then the integral in \TooMany\ would converge if
 \eqn{MuBound}{
  {\sqrt{g^{tt}_\infty} \bar\omega_\infty 
   \over \pi\mu\bar\omega(r)} < T_H \,.
 }
The left hand side acts in some rough sense like a temperature for the
background in question.  Similar results can be established for
singularities of $\bar\omega(t)^2$ in the complex $t$ plane: the
result is some coefficient other than $\pi$ in \MuBound.  We will
encounter such an alteration of \MuBound\ at the end of this section.

The type of result expressed in \MuBound\ provides a good intuitive
argument, albeit slightly circular, for why $\bar\omega(t)^2$ should
be analytic on the real line: zeroes and singularities of
$\bar\omega(t)^2$ in the complex plane have to be far enough away from
the real axis for \MuBound\ to be satisfied.  So the radius of
convergence of $\bar\omega(t)^2$ is finite everywhere on the real
line.

As an interesting class of examples, consider a $k=0$ FRW cosmology:
 \eqn{FRW}{
  ds^2 = a(\eta)^2 (-d\eta^2 + d\vec{x}^2) \,.
 }
The modes of a conformally coupled scalar with mass $m$ satisfy
\ChiEOM\ with $t$ replaced by $\eta$ and $\omega^2 = k^2 + m^2
a^2$.\footnote{Different choices of the parameter $\xi$ in the term
$\xi \chi^2 R$ that controls the coupling to the Ricci scalar result
in finite shifts in $\omega^2$.  As long as $\xi$ does not grow too
quickly as one goes to more highly excited states, it should not
affect the analysis at the level we are working at.}  Assuming that
the metric in \FRW\ is the string frame metric, and that the curvature
is sub-stringy, the mass spectrum is approximately given by $m^2 =
N/\alpha'$, where $N$ is the level.\footnote{Neither the zero point
nor the normalization of $N$ quite agrees with the conventional
definition of excitation level in type~II string theory.  This minor
discrepancy is of no consequence as long as we correctly keep track of
the normalization of $T_H$.}  For highly excited string states,
neglecting $k$ is a good approximation, though not a uniform one if
$a(\eta)$ becomes arbitrarily small in the far past.  Let us assume
that $a(\eta)$ approaches nonzero constant values, $a_{-\infty}$ or
$a_\infty$, as $\eta \to \pm\infty$.  The scaling analysis above is
appropriate, with $\bar\omega(\eta) = a(\eta)/\sqrt{\alpha'}$.
Assuming that a single zero $\eta_* = r-i\mu$ of $a(\eta)^2$
dominates, the condition \MuBound\ for there to be a finite number of
strings created is
 \eqn{FRWMany}{
  {1 \over \pi\mu a(r)} < T_H \,.
 }
Specializing further, let us consider one of the classic exactly
solvable problems:\footnote{This treatment is similar to the
one in \cite{ChungHeavy}.}
 \eqn{FRWsolvable}{
  a(\eta)^2 &= A + B \tanh(\rho\eta) \,,
 }
The exact result leads to exponentially suppressed particle production
for large masses:
 \eqn{SolvableBeta}{
  |\beta|^2 = {\sinh^2\left[
     \pi (\omega_\infty-\omega_{-\infty})/2\rho \right] \over
    \sinh(\pi\omega_\infty/\rho) \sinh(\pi\omega_{-\infty}/\rho)} 
   \approx e^{-2\pi\omega_{-\infty}/\rho} = 
    e^{-2\pi\sigma\bar\omega_{-\infty}/\rho} \,,
 }
where $\omega(\eta) = \sqrt{k^2 + a(\eta)^2 m^2}$ and
$\omega_{\mp\infty}$ are the limits of $\omega(\eta)$ in the far past
and future, and the approximate equality holds good in the limit where
$\omega_{\pm\infty} \gg \rho$.  The zero of $a(\eta)^2$ in the lower
half plane closest to the real axis is $\eta_* = -{i\pi \over 2\rho} +
{1 \over 2\rho} \log {A-B \over A+B}$, so from \MuApprox\ we obtain
 \eqn{ApproxBeta}{
  |\beta|^2 \approx e^{-c_1
    2\pi\sigma\bar\omega_{-\infty}/\rho} \qquad\hbox{where}\quad
   c_1 = {\pi \over 4} \sqrt{1 + B/A} \,.
 }
Because $0 \leq B/A \leq 1$, we have $\pi/4 \leq c_1 < \pi/\sqrt{8}$,
indicating fairly good agreement with the exact result (perfect
agreement would be $c_1=1$).  A criterion of the form \MuBound\ or
\FRWMany\ emerges immediately from estimating the total number of
strings produced from \TooMany\ and \SolvableBeta, only with a factor
of $c_1$ multiplying the left hand side.  Thus we conclude from this
example that the steepest descent method gives a reasonable
approximation of the criterion for finiteness of the number of strings
produced.

Let us conclude this section with a background that includes de Sitter
space in a certain limit, but has well-defined in and out regions away
from this limit.  The geometry is defined by \FRW\ with
 \eqn{NearlydSDef}{
  a(\eta)^2 = A + B {\eta \over \sqrt{\eta^2 + 1/\rho^2}} \,,
 }
where $A$, $B$, and $\rho$ are constants.  If $A>B>0$, this background
is qualitatively similar to \FRWsolvable, but in the limit where
$A=B>0$, the metric interpolates between the $k=0$ patch of $dS_4$ in
the far past and flat space in the far future.  To be precise, for
$A=B$ and $\eta \ll -\rho$, we have
 \eqn{NearlydS}{
  a(\eta)^2 \approx {L^2 \over \eta^2} (-dt^2 + d\vec{x}^2) \qquad
   \hbox{where}\quad L = \sqrt{A \over 2\rho^2} \,.
 }
Although the background defined by \FRW\ and \NearlydSDef\ is not
motivated by string theory considerations, it might be interesting as
a simple ``regulation'' of de Sitter space: one can for example define
an S-matrix for $A>B>0$ and then take the limit $A\to B$ to
investigate aspects of quantum field theory in de Sitter space.

The singularity of $a(\eta)^2$ in the lower half plane closest to the
real axis is a branch cut starting at $\eta_* = -i/\rho$.  (There is
also a zero at $\eta = -{i \over \rho} {A \over \sqrt{A^2-B^2}}$, but
this is always further from the real axis, and it disappears
altogether in the $A=B$ limit).  The condition \FRWMany\ becomes $T_s
< T_H$, where
 \eqn{TsFirst}{
  T_s = {1 \over \pi\mu a(r)} = {\rho \over \pi\sqrt{A}} \,.
 }
As remarked earlier, $T_s$ is an effective temperature for the
spacetime.  Interestingly, the estimate \TsFirst\ does not depend on
$B$ at all---though, clearly, $B$ must control a prefactor on the
total number of strings produced that vanishes as $B \to A$.  In this
limit, $T_s$ as estimated in \TsFirst\ differs by $\sqrt{2}$ from the
temperature of the de Sitter horizon in the far past, which is $T_{dS}
= 1/(2\pi L)$ with $L$ as defined in \NearlydS.  The discrepancy is
entirely due to the crudeness of estimating the integral in the second
factor of \SteepestApprox\ using an elliptical arc passing through
zero: evaluating that integral exactly for $A=B$ changes the estimate
\TsFirst\ so that $T_s = T_{dS}$ exactly.

It is tempting to interpret the divergence in \TotNum\ that arises
when \MuApprox\ is violated as dual to the development of an open
string tachyon stretched in the complex plane between the singularity
point $t_*$ and its complex conjugate $\bar{t}_*$.  Such an
interpretation has been offered in time-dependent backgrounds
resulting from well-defined stringy constructions
\cite{maldacenaEtAl}, and the idea of D-branes in imaginary time has
been further developed in \cite{gir}.

In general, one should consider the possibility that the singularities
of $\bar\omega(t)$ (or, more precisely, of the integrand in
\TotNumRescaled) in the complex $t$ plane are not pointlike or even
branch cuts, but cover finite regions of the plane.  This could
happen, for instance, if we replaced the $a(\eta)^2$ in \FRWsolvable\
by a continuous sum over different values of $\rho$.  A criterion like
\MuBound\ should still apply, where $\mu$ is roughly the closest
approach of the singular region to the real axis.

\section{Strings and preheating}
\label{Preheating}

A widely studied application of particle creation in cosmology is the
theory of preheating (see for example \cite{bt,kls,zhs}), whereby
coherent fluctuations of the inflaton $\phi$ around its minimum lead
to an oscillating mass term for another bosonic field, $\chi$, through
a term in the action proportional to $\phi^2 \chi^2$.  The resulting
variation in $\omega(t)^2$ can set up a parametric resonance, which
produces exponentially growing particle occupation numbers for $\chi$.
The variation of $\omega(t)^2$ cannot be expected to be perfectly
periodic, since the universe is expanding and there may be significant
deviations from pure quadratic behavior in the inflaton potential in
the region where the inflaton oscillates.  This limits the
amplification that parametric resonance can provide.  One must then
ask under what conditions parametric resonance is ``efficient,'' so
that an appreciable fraction of the energy stored in the inflaton
oscillations will be converted to $\chi$ particles via preheating.
Oversimplifying a little, the answer is that the mass of $\chi$ must
vary by a large factor: the maximum mass in each cycle of oscillation
must be many times greater than the minimum mass.  In such a
situation, there is a ``broad resonance,'' \cite{kls}, and a variety
of complicated effects like back-reaction and rescattering become
significant.  In \cite{zhs}, more precise analytic criteria were
developed to decide whether a resonance is efficient.  Let us explore
how the criteria for efficient extraction of energy from the inflaton
oscillations might be altered due to the presence of a Hagedorn
density of states.

Occupation numbers of excited string states should be small even if
strings do participate significantly in preheating.  The reason is
that resonances become exponentially narrow, and the rate of growth of
occupation numbers within a resonance becomes exponentially slow, as
the as one increases the tree-level mass.  There can still be a large
total number of strings produced because of the competing exponential
growth of the density of states.  To distinguish approximately where
the competing exponentials have equal ``strength,'' it is sufficient
to work with $|\beta|^2 \ll 1$ for all individual string modes, except
perhaps for those which are massless at tree level.  Our plan of
attack, then, is to use the methods developed in
section~\ref{Steepest} to study production of the massive excited
states.

It is very plausible that the effective string tension, as measured in
the four-dimensional Einstein frame, varies as the inflaton varies.
Let us as usual assume a $k=0$ FRW cosmology.  When the string frame
metric is weakly curved, the spectrum is given approximately by
$g_{\rm str}^{tt} \omega^2 = g_{\rm str}^{xx} k^2 + m^2 = g_{\rm
str}^{xx} k^2 + N/\alpha'$, where as usual $N$ is the excitation
level.  The four-dimensional Einstein metric is related to the
four-dimensional part of the string metric by a conformal
transformation: $ds_{4E}^2 = e^{\gamma\varphi/M_{\rm Pl,4}} ds_{\rm
str}^2$ where $M_{\rm Pl,4}$ is the four-dimensional Planck mass,
$\varphi$ is a canonically normalized scalar which is some combination
of the dilaton and the volume of the internal manifold, and $\gamma$
is some constant, presumably $O(1)$ in generic
compactifications.\footnote{The expression $ds_{4E}^2 =
e^{\gamma\varphi/M_{\rm Pl,4}} ds_{\rm str}^2$ is imprecise because
the combination of scalars in the exponent may not be an eigenvector
of the scalar kinetic operator in four-dimensional Einstein frame.
But the oversimplification will not matter in the subsequent
treatment.}  We may adjust the zero point of $\varphi$ so that
$\varphi=0$ at the minimum around which one oscillates during
reheating.  Evidently, $g_{4E}^{tt} \omega^2 = g_{4E}^{xx} k^2 +
e^{\gamma\varphi/M_{\rm Pl,4}} N/\alpha'$.  Using the usual ansatz
$ds_{4E}^2 = -dt^2 + a(t)^2 d\vec{x}^2$, we end up with
 \eqn{omegaOscillates}{
  \omega(t)^2 = k^2 a(t)^2 + e^{\gamma\varphi(t)/M_{\rm Pl,4}} 
   {N \over \alpha'} \,.
 }
Now, $\varphi$ is {\it not} necessarily the inflaton: there are many
scalars in four dimensions that specify the size and shape of the
internal manifold.  But it seems safe to assume, at least on grounds
of genericity, that there is some overlap between $\varphi$ and the
inflaton $\phi$.  If $\phi$ oscillates with frequency $\Omega$ between
slowly varying extremes $\varphi_{\rm min}$ and $\varphi_{\rm max}$,
then we may approximate \omegaOscillates\ by
 \eqn{OmegaApproximate}{
  \omega(t)^2 &= {\omega_+^2 + \omega_-^2 \over 2} + 
   {\omega_+^2 - \omega_-^2 \over 2} \cos \Omega t  \cr
  \omega_-^2 &= k^2 a(t)^2 + e^{\gamma\varphi_{\rm min}/M_{\rm Pl,4}}
   {N \over \alpha'}  \cr
  \omega_+^2 &= k^2 a(t)^2 + e^{\gamma\varphi_{\rm max}/M_{\rm Pl,4}}
   {N \over \alpha'} \,.
 }
Neglecting derivatives of slowly varying quantities like $\Omega$ and
$\omega_{\pm}$, the wave-function of a given bosonic string mode,
suitably scaled by some time-dependent factor, satisfies Mathieu's
equation,
 \eqn{Mathieu}{
  \left( {d^2 \over dz^2} + A - 2q \cos 2z \right) \chi = 0 \,.
 }
with $z = \Omega t/2$ and $A$ and $q$ specified by
 \eqn{MathieuParameters}{
  A \pm 2q = A_\pm = \left( {\omega_\pm \over 2\Omega} \right)^2 \,.
 }
For $(A,q)$ in special regions of the plane, solutions to \Mathieu\
exist whose average value grows with $z$ as $e^{\nu z}$, where $\nu$
is the imaginary part of the so-called Floquet exponent.  These
solutions are a manifestation of parametric resonance, but we will not
need them for the reasons reviewed above.  Instead, after the usual
``rescaling'' $\omega \to \sigma\bar\omega$, we find ourselves in the
situation described in equations \RescaledEOM-\MuBound.  In
\OmegaApproximate-\MathieuParameters, we have neglected the damping of
the inflaton oscillations due to expansion and to dissipation of
energy into the string states.  Including these effects will make the
first oscillation produce more strings than any of the subsequent
ones.  To be more precise: the zeroes of $\omega(t)^2 = \sigma^2
\bar\omega(t)^2$ as given in \OmegaApproximate\ are
 \eqn{tPoles}{
  t^{\pm}_{*,n} = {2\pi n \over \Omega} \pm {i \over \Omega}
   \log {\bar\omega_+ + \bar\omega_- \over 
     \bar\omega_+ - \bar\omega_-} \,,
 }
but we are going to assume that $t_* = t^-_{*,0} = r - i\mu$ makes the
biggest contribution to string production.  In order for the formal
expression \TotNumRescaled\ to converge, \MuBound\ must be satisfied,
which is to say
 \eqn{MuBoundAgain}{
  \bar\omega(r) \mu = {\bar\omega_- \over \Omega} 
   \log {\bar\omega_+ + \bar\omega_- \over \bar\omega_+ - \bar\omega_-} >
   {\sqrt{g^{tt}_\infty} \bar\omega_\infty \over \pi T_H} \,.
 }
The right hand side is a quantity of order unity.  We should regard it
as uncertain by at least a factor of $2$, given that
\OmegaApproximate\ is only a rough guess for $\omega(t)^2$, and the
factor of $1/\pi$ is only approximate, as the analysis of the example
\FRWsolvable\ showed.

If the bound \MuBoundAgain\ is satisfied with room to spare, then the
highly excited string states are hardly produced at all.  Conversely,
if it is violated, a huge number of excited string states are
produced, rendered finite, if need be, by a cutoff on the Hagedorn
spectrum.  A cutoff at an energy $\Lambda$ limits the number of
strings produced to something like $e^{\Lambda/T_H}$, which is very
large if $\Lambda$ is considerably higher than $T_H$.  So it seems
likely that the string production is more likely to be limited by the
available energy in the coherent inflaton oscillations.  Thus a
violation of \MuBoundAgain\ is very likely to mean that excited string
states will play a vital role in preheating, in the sense that a
significant fraction of the energy in the inflaton oscillations goes
into excited strings.  We obtained \MuBoundAgain\ from a single
oscillation of the inflaton.  Further oscillations of the inflaton are
then presumably rapidly damped, and even if a finite number of zeroes
of $\omega(t)^2$ contribute significantly to the total string
production, the result will probably only be to change a prefactor on
the expression for the total number of strings produced:
\MuBoundAgain, which comes from the exponent of \TooMany, is less
likely to be affected.\footnote{We might worry that parametric
resonance could alter \MuBoundAgain.  But the way that occupation
numbers $|\beta|^2$ significantly larger than \MuApprox\ arise in
parametric resonance is by having $|\alpha|^2$ significantly larger
than $1$, so that the approximation $\alpha(t) \approx 1$ that led to
\MasterBeta\ fails.  This is related to having exponential growth
$e^{\nu z} \approx e^{\nu\Omega t / 2}$ of certain mode functions.
Occupation numbers of order unity for excited string states should
result in a huge divergence in the total number of strings produced.
Exponentially suppressed occupation numbers for highly excited strings
means that the approximations that led to \MuBoundAgain\ should be
valid.

One can try to test this self-consistent reasoning further by adapting
the estimate of \cite{zhs} to a Hagedorn spectrum of bosonic states.
The result is $dN_{\rm tot}/dt \sim \int^\infty d\sigma \, \nu(\sigma)
e^{\sigma \bar\omega/T_H}$, where $\nu(\sigma)$ is the Floquet
exponent for a string state at level $\sigma^2$, as determined by the
$A$ and $q$ appearing in a rescaled version of \MathieuParameters.
This expression leads to agreement with \MuBoundAgain\ up to factors
of order unity, but it is hard to evaluate those factors.}

Neglecting the logarithm in \MuBoundAgain\ as well as various factors
of order unity leads to the approximate convergence condition
 \eqn{TensionIneq}{
  \sqrt{\tau_-} \gsim m_{\rm inflaton} \,,
 }
where $\tau_-$ is the minimum string tension as measured in
four-dimensional Einstein frame.  (The point here is that
$\bar\omega_- \propto \sqrt{\tau_-}$ while $\Omega$ is roughly the
inflaton mass at its minimum).  The criterion for there to be copious
string production due to coherent oscillations of the inflaton is that
the inequality \TensionIneq\ should be violated.  This happens in a
considerably broader regime of parameters than the regime that appears
in the standard analysis of when preheating.  In the standard
analysis, in order for the resonance to be ``efficient,'' the minimum
mass of the additional scalar $\chi$ must be many times smaller than
the inflaton mass, so that during many oscillations of the inflaton,
the parameters of the Mathieu equation fall in the range $2q < A < 2q
+ \sqrt{bq}$ for some constant $b$ of order unity.  In contrast, for
us, the analogous region of efficient preheating is $2q < A < 2xq$ for
some $x>1$.  Of course, from a purely quantum field theoretic
standpoint, it may seem contrived for the masses of an entire Hagedorn
spectrum to oscillate in unison---but that is where the perspective of
string theory makes natural things that otherwise would seem very
special.

Three potential problems with the idea that strings could have played
a significant role in preheating are as follows, in increasing order
of seriousness.  First, the Hagedorn spectrum might be cut off at a
sufficiently low scale to invalidate the above analysis.  Second, the
fundamental string tension is supposed to be near the Planck scale,
much higher than typical values for the inflaton mass, so despite the
considerations of the previous paragraph, it may still seem quite
unnatural to have \TensionIneq\ satisfied.  And third, super-heavy
stable particles could be produced in unacceptable numbers, some of
them with fractional charge.  Let us examine these issues more
closely.

 \begin{enumerate}
  \item Cutoff on the spectrum: Since we believe that the number of
degrees of freedom is thinned out dramatically by interactions at high
energies, we should introduce some energy cutoff $\Lambda$ and keep
only those states with energy less than $\Lambda$.  If $\Lambda$ is
only slightly above the string scale $1/\sqrt{\alpha'}$, of course all
of our arguments collapse, and \TotNum\ is not a good way at all to
estimate the number of string quanta created.  But in fact, there is
reason to think that $\Lambda \gg 1/\sqrt{\alpha'}$: for example, as
noted in \cite{LennyKol}, the typical size of a string is greater than
the ten-dimensional Schwarzschild radius associated with its total
mass until $N \sim 1/g_s^8$, where $N$ is the excitation level and
$g_s$ is the string coupling.  It would be consistent with the spirit
of holography to put $\Lambda$ at about the energy scale of strings at
this high excitation level: $\Lambda \sim 1/(\sqrt{\alpha'} g_s^4)$,
which is indeed a fairly high scale if $g_s$ is even moderately
small.\footnote{It is argued in \cite{LennyKol} that gravitational
self-interaction of strings compresses the spectrum and makes the
density of states rise faster than in the free theory, until at some
high scale string states might be in one-to-one correspondence with
the microstates of black holes.  Such a super-exponential increase in
the number of states one can actually produce in the way we have
discussed in this paper is disastrous, since almost no background will
have a finite total number of quanta produced.  Presumably this is
where a total reorganization of concepts is needed, perhaps with
holography playing a central role.}  Since the total energy that can
go into the strings is finite in preheating, but scales as an
exponential of the typical energy of the string modes produced, it
might be that total energy provides a process-dependent cutoff that is
often lower than $\Lambda$.

  \item Heaviness of fundamental strings: in conventional scenarios,
the square root of the string tension is roughly five orders of
magnitude larger than the inflaton mass ($10^{18} \, {\rm GeV}$ as
compared to $10^{13} \, {\rm GeV}$, roughly speaking).  In order to
achieve \TensionIneq\ without making the inflaton mass much bigger
than it is usually assumed to be, the string coupling needs to become
very small during the inflaton oscillations---or, more precisely, the
four-dimensional Planck scale, which scales as
 \eqn{MPlFour}{
  M_{\rm Pl,4} \sim (\Vol CY_3)^{1/2} g_s^{-1} \alpha'^{-2} \,,
 }
must attain values much larger than $1/\sqrt{\alpha'}$.  This is not
inconceivable, but it does seem contrived.  It has been pointed out to
me \cite{DvaliPrivate} that an inflaton mass during preheating
substantially larger than $10^{13} \, {\rm GeV}$ might be arranged in
string theory.  This lies outside the scope of the current
investigation.
 \item Production of super-heavy stable particles: Unless the
compactification manifold is simply connected (or unless open strings
can propagate throughout ten-dimensional spacetime), there will be
stable states corresponding to strings wound around non-contractible
cycles.  Such states can have fractional electric charge \cite{GSW},
and they would be very heavy, with mass $m \sim \ell/\alpha'$, where
$\ell$ is the length of the non-contractible cycle.  The trouble is
that there is a Hagedorn spectrum of excited strings with any given
winding number, and it is not clear that their production is
suppressed strongly enough compared to strings with no non-trivial
topology which can decay into string modes which are massless at tree
level.  Excited string states with winding number would relax to
stable, super-heavy particles.  The bounds on the density of such
particles are stringent because they would contribute to dark matter
density.  Tracing back the upper bound on dark matter density today to
the epoch of preheating leads to the conclusion that super-heavy stable
particles should comprise a very small fraction of total energy at
that time---perhaps on the order $10^{-17}$ \cite{ChungHeavy}.

We can estimate how rapidly such particles would be created as
follows.  The energy of states with non-trivial winding is
 \eqn{mKKwind}{
  \omega^2 = \omega_0^2 + m^2 \,,
 }
where $\omega_0^2$ is the energy of a string state with no winding,
but with the same momenta in the non-compact dimensions and at the
same excitation level.  The density of the states with winding should
be roughly the same, as a function of $\omega_0$, as the density of
states without winding.\footnote{This is easily seen to be so in the
case of a circle compactification when there is winding number but no
Kaluza-Klein momentum.  I doubt the Calabi-Yau case would be much
different, but I have not carried out the computations explicitly.}
Thus the calculations leading up to \TooMany\ can be adapted to an
approximate treatment of production of heavy charged particles by
replacing the density of states
 \eqn{ReplaceDensity}{
  e^{\omega/T_H} \longrightarrow 
   {\omega \over \sqrt{\omega^2-m^2}} e^{\sqrt{\omega^2-m^2}/T_H} \,.
 }
If production of highly excited string states dominates, then roughly
an equal number are produced with winding as without, simply because
$\sqrt{\omega^2-m^2} \approx \omega$ for these highly excited strings.
Prefactors on the density of states might suppress the production of
strings with winding by a few orders of magnitude---but suppression by
a factor of $10^{-17}$ seems difficult.  This is the relevant factor
because we are operating on the assumption that excited strings do
suck out a significant fraction of the energy of the coherent
oscillations of the inflaton field.  Naturally, if this assumption is
lifted, there is no pressing problem with overproduction of super-heavy
stable particles.\footnote{In fact, if we assume that the minimum
string tension (as measured in four-dimensional Einstein frame) is
about an order of magnitude higher than the inflaton mass, then the
factor $e^{-\pi\mu\omega(r)}$ suppressing string production (c.f.\
\MuApprox) can be arranged to be around $10^{-17}$.  This might
suggest a new twist on the idea of string-motivated super-heavy dark
matter.}

The considerably more stringent limits on fractionally charged
particles have not been invoked in the discussion of the previous two
paragraphs because it's not clear to me that fractionally charged
particles are a universal aspect of string models with non-trivial
topology in the extra dimensions.
 \end{enumerate}

In the analysis following \ReplaceDensity, we have assumed that string
production is not somehow limited to modes less energetic than
$\ell/\alpha'$.  Intuitively speaking, the point is that the spectrum
of strings with winding defines the same temperature (despite the
presence of a gap) as the strings without winding.  We have also
neglected the annihilation of strings wound in one direction with
strings wound in the other.  Some estimates of the cross-section might
be made, but annihilation processes don't seem likely to generate a
suppression factor anywhere close to $10^{-17}$.  Thus we regard the
overproduction of winding states as a major peril if highly excited
strings are assumed to play any role at all in
preheating.

\section{Non-perturbatively constructed strings}
\label{Light}

Points 2 and 3 from the end of the previous section may incline us to
doubt that excited strings can plausibly have anything to do with
preheating: on one hand, they're probably too heavy to be produced in
significant numbers, and on the other hand, if they are, there is a
tendency to overproduce super-heavy stable particles.

Strings in four dimensions arising from branes wrapped on cycles of an
internal manifold offer the possibility of ameliorating the problems
described above.  Regarding problem 2, the key point is that strings
coming from wrapped branes become tensionless at special points in the
moduli space which are a finite distance from generic points and
typically occur at real codimension two.  Regarding problem 3, the
helpful feature is that degenerating cycles can occur in an isolated
nearly singular region of the internal manifold, so that no state with
non-trivial topology is ever light.  We will expand on these points
somewhat in the following paragraphs and in the process do another
simple estimate of string production.

First, to see that wrapped branes must be included on an equal footing
with other states in the spectrum, recall for example the duality
between the heterotic string on $T^3$ and M-theory on K3, where when
cycles of the K3 shrink, wrapped M2-branes become light and provide
the W-bosons of enhanced gauge symmetry that can be seen by other
means on the heterotic side \cite{HullTownsend,WittenVarious}).

Arbitrarily light strings can arise in various ways
\cite{WittenComments,AndyOpen,ghSmall,swComments}.  A typical
situation is for the effective tension in four dimensions to be
 \eqn{tauEff}{
  \tau_{\rm eff} \approx M  |\varphi| 
   \qquad\hbox{with}\qquad M \sim {M_{\rm Pl,4} \over g_s} \,,
 }
where $\varphi$ is a canonically normalized complex scalar field.  The
precise dependence of $M$ on $g_s$ may depend on details, but some
inverse dependence on $g_s$ is to be expected from states that descend
from string solitons.  Let us consider D3-branes wrapped on a
shrinking $S^2$ within a Calabi-Yau manifold as a definite example.
The complex scalar $\varphi$ in this case is proportional to
$\int_{S^2} (J_2+iB_2)$.  The real part of this integral is the volume
of the $S^2$, and the imaginary part is an axion whose associated
instanton is a fundamental string wrapping the $S^2$.\footnote{There
are well-understood worldsheet instanton corrections to the metric on
the Kahler moduli space which we will be neglecting.  These
singularities are logarithmic and do not change the fact that the
point at which wrapped D3-branes become tensionless is at a finite
distance from generic points.  They may slightly alter \tauEff.}  The
action includes the standard DBI term,
 \eqn{DBIterm}{
  S = -\tau_{D3} \int d^4 \xi \, \sqrt{G_{\mu\nu} + B_{\mu\nu}} +
   \ldots \,.
 }
Integrating this on the $S^2$ and using $\tau_{D3} \sim 1/(\alpha'^2
g_s)$ gives the Nambu-Goto action for a string extended in other
directions, with a tension given by \tauEff.  The overall
normalization of the tension definitely depends on more data than
we've specified so far (for instance, the total volume of the
Calabi-Yau manifold), but \tauEff\ should be right up to factors of
order unity provided one is not in a peculiar corner of the moduli
space (such as a Calabi-Yau manifold which is many times larger than
the fundamental string scale).

An $S^2$ can indeed degenerate at a finite distance in moduli space in
a Calabi-Yau compactification, and in an isolated part of the
Calabi-Yau manifold.  A famous example is the resolved conifold
\cite{CandelasParkes,CandelasConifolds,StromingerConifolds}.  It is
also possible for degenerate spheres to arise along a two-dimensional
locus within the Calabi-Yau manifold: this locus might be a Riemann
surface of non-trivial topology, which resurrects the potential hazard
of overproduction of super-heavy stable particles.

At least in certain circumstances, it is clear that non-perturbatively
constructed strings have the same Hagedorn behavior as fundamental
strings.  The cleanest evidence comes from NS5-branes, where for many
coincident branes, the Hagedorn temperature of the fractional
instanton strings is exactly the Hawking temperature for the
near-extremal supergravity solution \cite{juanFive}; and from NCOS
strings, where the effective string tension of open strings is lowered
by applying an electric field, and the Hagedorn behavior is simple to
understand \cite{ggkrw}.  In both cases, $T_H \sim \sqrt{\tau_{\rm
eff}}$.  We will assume that this relation continues to hold, and that
the arguments that led to \TotNum\ still apply.

Assume now that the scalar $\varphi$ has some overlap with the
dilaton, and that it is undergoing coherent oscillations just after
the end of inflation.  It is unnatural to suppose that $\varphi$
passes exactly through zero, but it might reasonably be assumed to
pass near zero in the complex plane.  Perhaps naively, we will assume
that $\dot\varphi$ is nearly constant while $|\varphi|$ is small.  Let
$\varphi_0$ be the closest approach of $\varphi$ to the origin.  Then
the effective string tension varies with time like this:
 \eqn{tauVaries}{
  \tau_{\rm eff} \approx M |\varphi_0 + \dot\varphi t|
   = M \sqrt{|\varphi_0|^2 + (|\dot\varphi| t)^2} \,.
 }
Proceeding in the same spirit as in the previous sections, we suppose
that modes of the string in four dimensions have a dispersion
relationship $\omega(t)^2 = k^2 + N\tau_{\rm eff}$, where $N$ is
something like the level, and we are not trying to keep track of the
expansion of the universe during the brief moment when the effective
string tension becomes small.  The Hagedorn temperature is $T_H \sim
\sqrt{\tau_{\rm eff}}$.  The frequencies $\omega(t)$ (and also the
Hagedorn temperature) are slowly varying in the far future and the far
past, in the sense that derivatives of these quantities are much less
than the appropriate power of the undifferentiated quantities.  So
asymptotic in and out vacua can be defined, at least in the
approximation where we neglect what happens when the scalar rolls out
of the realm of validity of \tauEff, and particle production can be
estimated in the same way that led to \MuApprox.  If we define
$\bar\omega(t) = \sqrt{\tau_{\rm eff}(t)}$, then the result is that an
integral of the type found in \TooMany\ converges provided a criterion
similar to \MuBoundAgain\ holds.  Dropping various factors of order
unity, that convergence criterion is $\bar\omega(r)\mu \gsim 1$.  The
solution to $\tau_{\rm eff}(t) = 0$ in the lower half of the complex
plane is $t_* \equiv r-i\mu = -i |\varphi_0|/|\dot\varphi|$.  Thus the
convergence criterion is $\sqrt{M |\varphi_0|^3}/|\dot\varphi| \gsim
1$.  We can roughly estimate $\dot\varphi \sim M_{\rm Pl,4} m_{\rm
inflaton}$, on the assumption that the coherent oscillations of
$\varphi$ are of amplitude $M_{\rm Pl,4}$ and of frequency $m_{\rm
inflaton}$.  Recalling that the minimum tension is $\tau_{\rm eff,-}
\approx M|\varphi_0|$, we can summarize the convergence criterion as
follows:
 \eqn{BoundSummary}{
  & 1 \lsim \bar\omega(r) \mu = {\sqrt{M \varphi_0^3} \over \dot\varphi}
   = {\tau_{\rm eff,-}^{3/2} \over M \dot\varphi}
   \sim {\tau_{\rm eff,-}^{3/2} \over M M_{\rm Pl,4} m_{\rm inflaton}}
   \sim 10^4 \left( {\tau_{\rm eff,-} \over M_{\rm Pl,4}^2} \right)^{3/2}
    \cr
  & \hbox{or equivalently,}\quad 
   \sqrt{\tau_{\rm eff,-}} \gsim {1 \over 20} M_{\rm Pl,4} \,,
 }
where we have assumed $M \sim 10 M_{\rm Pl,4}$ and $m_{\rm inflaton}
\sim 10^{-5} M_{\rm Pl,4}$.  The estimates leading to \BoundSummary\
are inevitably fairly crude as long as we have not specified what the
inflaton is and how much it overlaps with the scalar $\varphi$
controlling the effective string tension.  The result in
\BoundSummary\ is rather different from \MuBoundAgain\ and
\TensionIneq.  Intuitively, this is because the harmonic dependence
that led to \MuBoundAgain\ is the smoothest interpolation between a
given maximum and minimum in a specified time interval, whereas
\tauVaries, though still analytic, implies almost a discontinuity in
the first derivative of $\omega(t)^2$.

Evidently, the non-perturbatively constructed strings do not have to
dip very far below the Planck scale in order to violate the
convergence criterion \BoundSummary.  As before, the exponential
nature of the divergence then would seem likely to be efficient in
sucking energy out of the coherent motion of the inflaton.  More
precisely, it would be efficient in sucking out the energy from the
coherent motion of $\varphi$, which is probably not the only component
of the inflaton.  This might be modeled in a Hartree approximation by
a damping term for the evolution of one scalar which acts only near a
certain value of that scalar, while other scalars, coupled to the
first by terms in the potential, are only damped by the usual
$3H\dot\phi$ term.

\section{Conclusions}
\label{Conclude}

The convergence criterion on the total number of strings produced,
which we have variously stated as \MuBound, \MuBoundAgain, and
\BoundSummary, amounts to the statement that in the appropriate string
frame, where the string tension is by definition constant, curvatures
are substringy---only it is crucial that not only the second
derivatives of the string metric are bounded by the string scale, but
all derivatives are bounded in an appropriate way to ensure a uniform
radius of convergence, of order the string length, for the
time-varying energies of string modes.  In the context of preheating,
a violation of this bound on curvatures, either for fundamental
strings or for strings constructed as wrapped branes, has the simple
interpretation that the production of highly excited string states
efficiently drains the energy of coherent inflaton oscillations.  This
is somewhat in analogy with the standard theory of preheating, but
copious string production occurs in a larger region of parameters (as
measured in the $(A,q)$ plane of parameters for Mathieu functions).
Unfortunately, it is not particularly clear to me what distinctive
experimental signatures should be expected if highly excited strings
played a significant role in preheating.  The strings decay into
states that are massless at tree level, and presumably these can then
thermalize.  However, if excited strings states are created in
appreciable numbers, it seems that stable super-heavy particles are
likely to be overproduced, if they exist in the string spectrum.  This
problem can be avoided by considering strings in four dimensions that
arise as branes wrapped on cycles of an internal manifold---cycles
which shrink to produce a point-like singularity on that manifold.

The tension of such non-perturbatively constructed strings vanishes at
certain points in moduli space, and near such points the dependence of
the tension on canonically normalized scalars may reasonably be
assumed to be \tauEff.  This dependence leads to a convergence
criterion \BoundSummary\ which is considerably more restrictive than
the one derived in \MuBoundAgain\ and \TensionIneq\ assuming harmonic
dependence---meaning that the minimum tension doesn't have to be as
low, assuming the dependence \tauEff, in order to violate the
convergence criteria and produce a large number of excited strings.
In fact, as we see in \BoundSummary, the effective string tension only
needs to dip a couple of orders of magnitude below the
four-dimensional Planck scale in order to have copious string
production, with otherwise rather standard assumptions about the
amplitude and frequency of inflaton oscillations just after the end of
inflation.

In summary, it is quite plausible that excited strings (probably of
non-perturbative origin) play a role in post-inflationary cosmology.

A much-studied scenario in which light strings appear is ekpyrosis
\cite{kost,kosst}, in which either an M5-brane wrapped on a
holomorphic cycle of a Calabi-Yau manifold collides with the $E_8$
boundary of spacetime constructed in \cite{hw}, or the two $E_8$
boundaries collide with one another.  In the first case, M2-branes
stretched between the M5-brane and the boundary behave as strings that
become light at the moment of collision.  In the second case,
M2-branes stretched between the two $E_8$ boundaries become light, and
they become precisely the perturbative heterotic strings in the limit
where the two boundaries coincide.  The results of this paper bear on
ekpyrosis in two ways:
 \begin{enumerate}
  \item We have noted in section~\ref{Light} a natural way of
avoiding or ``regulating'' the collision: one simply needs to assume a
non-zero value for the pseudoscalar superpartner of the real scalar
controlling the distance between the branes/boundaries.  In the case
of colliding boundaries, this peudoscalar is just the axion arising
from the Neveu-Schwarz $B_{\mu\nu}$ field of the heterotic string with
both indices in four dimensions.  If this axion exists in the
spectrum, it is in fact unnatural to suppose that it is zero during
ekpyrosis.
  \item If ekpyrosis does occur, presumably in a ``regulated'' way as
explained in the previous point, then estimates of the creation of
light strings proceed in parallel to our calculations in
section~\ref{Light}.  If excited strings are produced during
ekpyrosis, the problem of overproducing stable super-heavy particles
seems likely to recur.  Specifically, if an M5-brane wrapped on a
holomorphic curve hits an $E_8$ boundary, then the light strings can
wrap cycles of the holomorphic curve (which is a Riemann surface
embedded in the Calabi-Yau manifold), unless that curve is
topologically an $S^2$.  And if the two boundaries collide, then the
light strings can wrap on the Calabi-Yau, unless the Calabi-Yau is
simply connected, which would considerably reduce the
particle-phenomenological appeal of the setup.

It doesn't help to assume that the strings are light only for a brief
instant of time.  In fact, according to \BoundSummary, larger
$\dot\varphi$ (meaning a briefer collision) causes more string
production, not less: it's then easier to violate the inequality in
\BoundSummary.
 \end{enumerate}

The reason that it is easy to offer the plausible interpretation of
draining energy from coherent inflaton oscillations when strings are
copiously produced just after inflation is that a significant fraction
of the energy is kinetic (relating to the time-derivative of the
inflaton).  A much thornier question is what happens if a string
becomes lighter than the Hubble scale during slow-roll inflation, when
most of the energy is stored in the scalar potential.  I have
speculated elsewhere \cite{gubserTalk} that light strings might limit
the rate of inflation to the string scale, and that there could be a
fixed point of forward time evolution where a string tension is
vanishing at the same time that the expansion of the universe is
slowing to something less fast than exponential.  An observation that
tells against this idea is that string production should result in a
new source of positive energy density, and unlike the case of
preheating, there is no obvious way to suck this energy away from the
scalar potential energy.  Granting the Friedman equation $H^2 =
\rho/(3 M_{\rm Pl,4}^2)$, it is hard to see what mechanism due to
light strings will drive $\rho$ and therefore $H$ to zero.  Perhaps
this question must wait on some treatment more firmly based on the
first principles of string theory.

Indeed, an objection that could be raised to the entire enterprise of
this paper is that $\alpha'$ corrections to the spacetime equations of
motion become significant when curvatures are string scale, so it's
not clear that we have adequately addressed classical effects before
delving into the quantum effect of particle creation.  There are at
least three reasons to continue thinking about string creation in
general backgrounds: 1) Non-perturbative strings don't have a
well-understood connection to spacetime equations of motion.  2)
Worldsheet techniques have not so far provided a particularly large
class of well-understood time-dependent backgrounds.  3) Because
copious string creation occurs, in some heuristic sense, at the same
order in $\alpha'$ as corrections to the classical spacetime
equations, string production probably does have an important role to
play in our understanding of general time-dependent backgrounds.

Let us end with some open questions:
 \begin{itemize}
  \item Our estimates for string production were fairly crude in
places, starting with the neglect of power law corrections to the
exponential growth in the density of states, and including also the
neglect of various factors of order unity that enter into the
exponential behavior in \TooMany.  Is it possible to give more precise
estimates?
  \item Can one give an explicit string theory construction of
time-dependent backgrounds in which highly excited string states are
produced?  This is surely complicated by the fact that back-reaction
from such strings is likely to be important.
  \item Might the CP violation that results from turning on an axion
to ``get around'' a point of tensionless strings be of interest in
baryogenesis?
  \item If highly excited strings are produced when inflation is over
or nearly over, is there some definite affect on the spectrum of
fluctuations?  How do decay products of highly excited strings
thermalize?
  \item If we assume that the end of inflation involves oscillations
in a complex moduli space with tensionless strings on some real
codimension two loci, how much isocurvature perturbations are likely
to be generated?  Is the scenario plausibly consistent with existing
bounds?
 \end{itemize}
We hope to return to some of these questions in the future.

\section*{Acknowledgments}

I thank J.~McGreevy for reminding me of the origin of the heterotic
string axion, H.~Verlinde for discussions regarding light strings and
ekpyrosis, and C.~Callan, D.~Freedman, S.~Shenker, and P.~Steinhardt
for other useful discussions.  I am particularly grateful to G.~Dvali
for discussions on various possible effects of light strings on
cosmology.  This work was supported in part by the Department of
Energy under Grant No.\ DE-FG02-91ER40671.

\newpage
\bibliography{contour}
\bibliographystyle{ssg}

\end{document}